\documentclass[showpacs,floats,twocolumn,floatfix,superscriptaddress]{revtex4-1}

\usepackage{subfigure}
\usepackage{graphics,graphicx}
\usepackage{amssymb,amsmath}
\usepackage[utf8]{inputenc}
\usepackage[normalem]{ulem}

\renewcommand{\vec}[1]{\mathbf{#1}}

\usepackage[dvipsnames]{xcolor}

\begin{document}

\title{Mesoscale simulation of soft particles with tunable contact angle in multi-component fluids}

\author{Maarten Wouters}
\email{m.p.j.wouters@tue.nl}
\affiliation{Department of Applied Physics, Eindhoven University of Technology, De Rondom 70, 5612 AP, Eindhoven, The Netherlands}

\author{Othmane Aouane}
\email{o.aouane@fz-juelich.de}
\affiliation{Helmholtz Institute Erlangen-N{\"u}rnberg for Renewable Energy, Forschungszentrum J{\"u}lich, F{\"u}rther Strasse 248, N{\"u}rnberg, Germany}

\author{Timm Kr\"uger}
\email{timm.krueger@ed.ac.uk}
\affiliation{School of Engineering, Institute for Multiscale Thermofluids, The University of Edinburgh, Edinburgh EH9 3FB, Scotland, UK}

\author{Jens Harting}
\email{j.harting@fz-juelich.de}
\affiliation{Helmholtz Institute Erlangen-N{\"u}rnberg for Renewable Energy, Forschungszentrum J{\"u}lich, F{\"u}rther Strasse 248, N{\"u}rnberg, Germany}
\affiliation{Department of Applied Physics, Eindhoven University of Technology, De Rondom 70, 5612 AP, Eindhoven, the Netherlands}


\begin{abstract}
Soft particles at fluid interfaces play an important role in many aspects of
our daily life, such as the food industry, paints and coatings, and medical
applications.  Analytical methods are not capable of describing the emergent
effects of the complex dynamics of suspensions of many soft particles, whereas
experiments typically either only capture bulk properties or require invasive
methods.  Computational methods are therefore a great tool to complement
experimental work.  However, an efficient and versatile numerical method is
needed to model dense suspensions of many soft particles.  In this article we
propose a method to simulate soft particles in a multi-component fluid, both at
and near fluid-fluid interfaces, based on the lattice Boltzmann method, and
characterize the error stemming from the fluid-structure coupling for the
particle equilibrium shape when adsorbed onto a fluid-fluid interface.
Furthermore, we characterize the influence of the preferential contact angle of
the particle surface and the particle softness on the vertical displacement of
the center of mass relative to the fluid interface.  Finally, we demonstrate
the capability of our model by simulating a soft capsule adsorbing onto a fluid-fluid interface with a shear flow parallel to the interface, and
the covering of a droplet suspended in another fluid by soft particles with different wettability.
\end{abstract}


\maketitle

\section{Introduction}
In many fields soft particles at interfaces play an important role, for example
medical applications~\cite{Levy1990,Constantinides1995},
cosmetics~\cite{Puglia2012}, food industry~\cite{Silva2012}, and paints and
coatings~\cite{Kooij2016,Keddie2010}. For a single particle one can still
analytically describe simple systems.  However, when multiple particles
interact with each other, it is typically no longer possible to predict the
deformation dynamics of the individual particles, especially when fluid-fluid
or fluid-vapor interfaces are present.

Computational methods can be of great value to complement experiments and to
characterize emergent phenomena stemming from the complex particle-particle and
particle-fluid interactions, such as the margination of platelets in
blood-flow~\cite{Krueger2016,Mehrabi2016}, the rheology of a suspension of soft
capsules~\cite{Gross2014,MacMeccan2009,Krueger2014}, or the stabilization of emulsions with
(solid) nano-particles~\cite{Frijters2012,Gunther2014}.

Simulating soft particles at fluid interfaces can be done on the microscopic
scale, with for instance molecular dynamics simulations where one resolves the
polymer chains or molecules which make up the soft
particle~\cite{Mehrabian2016}. Such methods are, however, extremely costly, and
therefore limited to simulations of only a few particles. Macroscopic methods
typically directly solve the Navier-Stokes equations, or a coupled set of
equations, discretized onto a grid to make it numerically accessible, such as
for example finite element methods. To properly describe fluid-structure interactions this
often requires expensive re-meshing steps and thereby becomes increasingly
complex to solve, both in implementation and computational
effort~\cite{Zienkiewicz2014}.

Our goal is an efficient method which makes large-scale simulations of
suspensions of soft particles feasible on existing high-performance computing
clusters, and offers the possibility to easily extend the simulated system to
more complex cases.
The mesoscopic lattice Boltzmann method (LBM) is a suitable candidate
since it is straightforward to implement and has proven its capability of
efficiently simulating large systems of suspended particles in a
fluid~\cite{Ladd2001,Aidun2010,Harting2004}.

Simulations of soft particles in a single-component fluid are well established
in the LBM literature. Typical examples are fluid-filled capsules and
cells~\cite{MacMeccan2009,Krueger2011a,Gross2014,Gekle2017},
vesicles~\cite{Kaoui11,Farutin2018,Halliday2016}, and particles with
three-dimensional elasticity~\cite{Buxton2001,Wu2010_2}. Comparably, there are
various approaches to model rigid particles in multi-component fluids. Several
authors apply the momentum exchange method introduced by
Ladd~\cite{Ladd1994_1,Ladd2001} for the solid particle, coupled to the
pseudo-potential multi-component fluid method of Shan and
Chen~\cite{Joshi2009,Jansen2011,Frijters2014,Cheng2017} or a free-energy based
multi-component model~\cite{Stratford2005,Connington2015}. Other methods for
the fluid-structure coupling exist as well, such as fluid particle
dynamics~\cite{Tanaka2000}, or the immersed boundary
method~\cite{Feng2004,Wu2009}.

Despite the abundance of methods for rigid colloids in multi-component fluids,
the simulation of soft particles in multi-component fluids is still sparse in
the LBM literature.  Li et al.~\cite{Li2016} coupled the pseudo-potential
method of Shan and Chen~\cite{Shan1993,Shan1995} with the immersed boundary
method of Peskin~\cite{Peskin1972}.
Pepona et al.~\cite{Pepona2019} have recently presented a two-dimensional model for
soft particles at fluid-fluid interfaces based on the free-energy LBM coupled with
immersed boundaries.
The immersed boundary method interpolates
the local velocity around each boundary node via an interpolation stencil in
order to acquire the velocity of the boundary node.  Overlap of these
interpolation stencils is inherent to the model, resulting in a small mass-flux
through the boundary.  Li used an improved scheme which corrects for the
overlap of the stencils in order to ensure the no-slip condition at moving
solid boundaries~\cite{Pinelli2010}, and finally applied the method to
deformable cilia, which are partially at the interface between two fluids.
Their method is however costly due to the need of inverting a square matrix of
the size of the number of discretization points for the boundary.  Since their
cilia are discretized with roughly 20 boundary nodes, this does not pose a
significant problem for the computational efficiency.  However, for systems
with many soft particles in which a single particle can typically have of the
order of $\mathcal{O}(10^3)$ boundary nodes, the impact of this matrix
inversion quickly becomes a computational bottleneck.

In this work we propose an alternative approach which is capable of describing
soft particles with a tunable contact angle, in three dimensional
multi-component fluids in large and/or dense systems. It combines several
existing methods commonly used in the LBM literature.  Our basis is the
momentum exchange bounce-back fluid-structure coupling as introduced by
Ladd~\cite{Ladd1994_1,Ladd2001}, coupled to soft particles in a similar manner
as introduced for single-component systems by Aidun et
al.~\cite{Reasor2011,MacMeccan2009,Cheng2017,Aidun2010}. This is then coupled
to the popular multi-component pseudo-potential method of Shan and Chen
(SC)~\cite{Shan1993,Shan1995} and allows for tunable preferential contact
angles of the particle surface via an effective off-set in the densities used
for the force calculations as introduced by Jansen et al.~\cite{Jansen2011}.

The presented method has the benefit that the no-mass-flux condition across the
particle boundary is guaranteed inherently thanks to the bounce-back
conditions. Furthermore, the fluid-structure coupling is independent of the
collision step of the LBM and the chosen constitutive models for the soft
particles, which ensures a lot of freedom for future extensions or adaptations
of the model. Finally, our method allows to tune properties at the level of
the used discretization of the boundary, opening the possibility to for example
simulate soft particles with inhomogeneous wetting properties across the
surface. 

For clarity, we present the method in the most simple implementation possible and refer to the existing literature for possible extensions and improvements.
We use the Skalak (SK) hyperelastic constitutive law~\cite{Skalak1973} to model
the fluid-filled capsules throughout this work, but we emphasize that other
existing soft particle models~\cite{Barthes2016,Biben2011,Buxton2001} can be
applied straightforwardly as well.

The remainder of the paper is structured as follows. In
section~\ref{sec:method} we introduce the simulation method and provide brief
summaries of the different existing models combined in our approach.  We
validate our method in section~\ref{sec:results}. First, the deformation of a
single capsule in shear flow is compared to previously reported values in the
literature.  Then, the error in the equilibrium shape of capsules adsorbed onto a
fluid-fluid interfaces is studied, where we compare our data to a reference
model to explicitly exclude all causes of errors other than the used
fluid-structure coupling, since the used models for soft particles are already
well reported in literature. Also the influence of a shear flow parallel
to the fluid-fluid interface on the steady-state shape of a soft particle is briefly studied.
Furthermore, we demonstrate the possibilities of
our method by simulating the adsorption of soft capsules with different
wettability onto a droplet. Finally, we summarize and discuss our results in
section~\ref{sec:conclusion}.

\section{Method}\label{sec:method}

\subsection{The lattice Boltzmann method}\label{sec:method:lbm}
To simulate the dynamics of the fluids, we apply the lattice Boltzmann method
(LBM).  It allows for a straightforward implementation and has proven its
capability to handle a wide variety of fluid dynamics
problems~\cite{Benzi1992,Succi2001,Bible}. Thanks to the locality of the algorithm it is
also well suited for a parallel implementation and high performance computing
applications~\cite{Aidun2010,Harting2004}.

The LBM is based on the discretization of the Boltzmann equation, where we
discretize space into a regular 3D lattice with positions $\vec{x}$ and lattice
spacing $\Delta x$, and the time $t$ into evenly spaced time steps with step
size $\Delta t$. Additionally, we discretize the velocity space 
with 19 discrete lattice velocities $\vec{c}_i$ (D3Q19 model).  Using the
Bhatnagar-Gross-Krook (BGK) collision operator~\cite{Bhatnagar1954} the time
evolution of the single-particle probability distribution function $f_i$ is
described by
\begin{multline}
    f_i(\vec{x}+\vec{c}_i\Delta t, t+\Delta t) - f_i(\vec{x}, t) =  \\
        -\frac{\Delta t}{\tau}\bigg[f_i(\vec{x},t) - f_i^{eq}(\vec{x},t) \bigg],
    \label{eq:method:boltzmann}
\end{multline}
where $\tau$ is a relaxation time associated with the kinematic viscosity as
$\nu =c_s^{2}(\tau -0.5)\Delta x^2/\Delta t$, and $c_s=\sqrt{1/3}\Delta
x/\Delta t$ is the speed of sound.

The left-hand side of Eq.~\ref{eq:method:boltzmann} describes the
streaming step during which the distribution functions $f_i$ are
streamed from fluid node $\vec{x}$ along their respective lattice directions
$\vec{c}_i$ to the neighboring fluid node $\vec{x}+\vec{c}_i$.  The right hand
side denotes the inter-molecular collisions and resulting relaxation towards a
local equilibrium distribution with characteristic timescale $\tau$.  We use a
second order accurate approximation for the equilibrium distribution
$f_i^{eq}$, given as 
\begin{equation}
    f_i^{eq}(\vec{x},t) = w_i \rho \bigg[1 + \frac{1}{c_s^2}(\vec{c}_i\cdot\vec{u}) + \frac{1}{2c_s^4}(\vec{c}_i\cdot\vec{u})^2 - \frac{1}{2c_s^2}u^2\bigg],
    \label{eq:lbm:equilibrium}
\end{equation}
where the lattice weights $w_i$ are defined as
\begin{equation}
    w_i = \begin{cases}
    1 / 3  & i = 0\hspace{1.2cm} \textnormal{(resting)}, \\
    1 / 18 & i = 1\ldots6\hspace{5mm}\textnormal{(orthogonal)}, \\
    1 / 36 & i = 7\ldots18\hspace{3.5mm}\textnormal{(diagonal)}
    \end{cases}
    \label{eq:lbm:weights}
\end{equation}
to ensure isotropy.

We include forces $\vec{F}(\vec{x})$ acting on the fluid as a shift in the
equilibrium velocity in Eq.~\ref{eq:lbm:equilibrium}~\cite{Shan1993}.  From the
distribution function $f_i$ the local macroscopic density $\rho$ and velocity
$\vec{u}$ can now be computed as
\begin{align}
    \rho(\vec{x}) &= \rho_0 \sum_i f_i(\vec{x}), \\
    \rho(\vec{x})\vec{u}(\vec{x}) &= \rho_0 \sum_i f_i(\vec{x}) \vec{c}_i + \frac{\rho_0\Delta t}{2} {\vec{F}(\vec{x})},
\end{align}
where $\rho_0$ is an arbitrary scaling factor for the mass density of the fluid, which is fixed to unity throughout this work.
For more details and an extensive derivation of the LBM we refer the interested reader to~\cite{Bible,Succi2001}.

\subsection{Soft particle model}\label{sec:method:elastic}
We consider a fluid-filled capsule modeled as a two-dimensional incompressible hyperelastic membrane discretized into regular triangular finite elements. The membrane mechanical properties depend on the chosen hyperelastic law. We use in this work the Skalak constitutive model which provides the membrane with resistance to local area dilatation and shear elasticity ~\cite{Skalak1973}. The corresponding strain energy reads as 
\begin{equation}
    E^\textnormal{strain} = \frac{\kappa_S}{4} \oint (I_1^2 +2I_1 - 2I_2 + C I_2^2) dA, \quad C > -1/2
    \label{eq:method:strain}
\end{equation}
where $\kappa_S$ is the shear elastic modulus, $I_1=\lambda_1^2+\lambda_2^2-2$ and $I_2=\lambda_1^2\lambda_2^2-1$ are the strain invariants of the deformation tensor, and $\lambda_1$ and $\lambda_2$ are the principal stretching ratios. $C$ is a parameter controlling the local inextensibility of the membrane and is related to the area dilation modulus by $\kappa_A=(1+2C)\kappa_S$. In the asymptotic limit of small deformation, the 2D Poisson ratio can be expressed as function of $C$ such as $\nu_s=C/(1+C)$ with $\nu_s \in~]-1..1]$~\cite{Barthes2016}. A constraint on the volume variation is enforced using a penalty function given by
\begin{equation}
    E^\textnormal{volume} = \frac{\kappa_V}{2} \frac{(V - V^0)^2}{V^0},
    \label{eq:method:volume}
\end{equation}
with $\kappa_V$ a volume-change resistance modulus, and $V^0$ the reference volume of the stress-free particle. The strain and volume forces are computed from their respective energy terms via the principle of virtual work~\cite{Charrier1989,Shrivastava1993}, where the force for each individual boundary node $\vec{x}_b$ is computed using
\begin{equation}
    \vec{F}_b = -\frac{\partial E(\vec{x}_b)}{\partial\vec{x}_b}.
\end{equation}
The membrane can be endowed with a bending resistance to out of plane deformation by means of the Helfrich free energy~\cite{Helfrich1973} given by
\begin{equation}
    {E}^\textnormal{bending} = \frac{\kappa_B}{2} \oint (2H - H_0)^2 dA,
    \label{eq:method:curvature}
\end{equation}
where $\kappa_B$ is the bending modulus, $H$ and $H_0$ are the mean and spontaneous curvatures, respectively. For the sake of simplicity the spontaneous curvature is neglected in this study. The bending surface force density is obtained from the functional derivative of Eq.~\ref{eq:method:curvature} and reads as
\begin{equation}
    \vec{f}_B = 2 \kappa_B \bigg[2H (H^2 - K) + \Delta_s H\bigg]\vec{n},
    \label{eq:surf_dens}
\end{equation}
where $K$ is the Gaussian curvature, $\Delta_s$ is the Laplace-Beltrami
operator, and $\vec{n}$ is the normal vector pointing outward from the particle
surface. The nodal bending force is obtained by multiplying
Eq.~\ref{eq:surf_dens} with the corresponding nodal area (Voronoi area). $H$,
$K$, and $\Delta_s$ are computed using the discrete differential-geometry
operators approach as detailed in Meyer et al.~\cite{Meyer2003}. We define the
dimensionless bending parameter $B=\kappa_B/(\kappa_S R_0^2)$, and fix it
throughout this work to $B=10^{-2}$ unless specified otherwise.

We apply a first-order accurate forward Euler scheme for the movement of the
boundary nodes, to prevent the costly need to re-calculate all the forces
acting on the boundary nodes at a higher rate than the LBM updates, as it would be
required by sub-step integration techniques.
Higher order time-integration schemes can be used as well, but were found
unnecessary for the presented work.

The proposed simulation method of this work can easily be applied to different
types of soft particles, such as fluid-filled elastic
capsules~\cite{Barthes2002,Barthes2016}, vesicles~\cite{Biben2011}, or soft
elastic particles with 3D elasticity~\cite{Buxton2001}. For the latter 3D
elastic particles one actually introduces a cross-linked lattice of boundary
nodes, rather than a 2D membrane.

\subsection{Fluid-structure coupling}\label{sec:method:bounceback}
Exact boundary conditions, which prevent any mass-flux through the particle
surface, are desirable for a correct physical model.  For single-component
fluids a small mass-flux is however typically mitigated by the pressure, where
the results can be considered tolerable based on the required accuracy.  For a
multi-component fluid the requirement of exact boundary conditions is more
important, since the interior and exterior fluid of the particle typically have
a different composition, where the mass-flux through its surface could for
example trigger the formation of droplets.  Hence, we ensure that no mass moves
through the surface by applying bounce-back boundary conditions for
distribution functions $f_i$ that cross a boundary element during the streaming
step.  We use the half-way bounce-back scheme (BB), as already used for the
simulation of soft particle suspensions in single-component
fluids~\cite{MacMeccan2009,Reasor2011,Clausen2010}, which bounces the incoming
distribution functions from a fluid node $\vec{x}_f$ onto a boundary element
back during the streaming step as~\cite{Ladd1994_1}
\begin{equation}
    f_j(\vec{x}_f, t+\Delta t) = f_i^{*}(\vec{x}_f, t) + 2\rho w_i \vec{c}_i\cdot\vec{v}_0 / c_s^2,
    \label{eq:method:SBB}
\end{equation}
where the superscript $^*$ indicates the post-collision state, $\vec{c}_i = - \vec{c}_j$, $\rho$ is taken as the local density at fluid node $\vec{x}_f$, and $\vec{v}_0$ is the velocity of the boundary element.

The momentum transfer onto the boundary is given by the difference of the
incoming and outgoing population as
\begin{equation}
    \Delta\vec{p}_i = \Delta x^3 \bigg[ f_i^*(\vec{x}_f,t) + f_j(\vec{x}_f, t+\Delta t)\bigg]\vec{c}_i.
    \label{eq:method:momentumtransfer}
\end{equation}

In the presented work the coupling of quantities between the boundary elements and boundary nodes 
is done via a homogeneous scheme where the three nodes of only the corresponding boundary element are given the same weighting factor.
This technique maintains the locality of the algorithm by eliminating the need to communicate properties of the boundary node of the neighbouring boundary elements.
Alternatively, both Buxton et al.~\cite{Buxton2005} and MacMeccan et al.~\cite{MacMeccan2009} introduced different weighting schemes for the spreading of the exchanged momentum to boundary nodes for all nodes within a characteristic distance.
However, no differences were observed on the time scale of equilibration for all presented simulations.
Hence, for the sake of brevity a thorough characterisation of this free parameter is left for future work. Also the interplay with the error stemming from the half-way bounce-back as compared to more accurate methods should be included in such a study.

The half-way bounce back method as given in Eq.~\ref{eq:method:SBB} is known to suffer from staggered momenta~\cite{Ladd1994_1,Ginzbourg1996,Khirevich2015}.
Especially in small enclosed volumes, such as our fluid-filled capsule, these momenta rapidly manifest themselves as a staggering in the velocity field.
In this work we prevent this by spreading the total exchanged momentum from the half-way bounce-back coupling homogeneously over two consecutive time steps~\cite{Ladd1994_1}.

The staggered momenta can also be prevented by using full-way
bounce-back~\cite{Sukop2006}, where the boundary is still approximately
half-way between two fluid nodes, but the populations $f_i$ travel the complete
distance between two fluid nodes and are reversed during the next collision
step.  However, for the full-way bounce-back one needs to store the populations
that pass a boundary element, which is impracticable when there are fluid nodes
on both sides of the boundary.  Alternatively, one could also increase the
accuracy of the discretization of the boundary with interpolated bounce-back
boundary conditions~\cite{Bouzidi2001}, multi-reflection boundary
conditions~\cite{Ginzburg2003}, or the one-point second-order bounce-back boundary condition~\cite{Tao2018}.
They do offer a better representation of the
boundary on the fluid lattice, and remove the force fluctuations that stem from
the jump that the location of the discretized boundary makes when a boundary
element moves past a fluid node in the half-way bounce-back algorithm.  However,
they are computationally more expensive since one needs to calculate the exact
distance between a fluid node and the boundary element along the lattice
vectors.  Furthermore, the interpolated bounce-back and multi-reflection boundary conditions
use information of multiple fluid nodes, which
increases the minimum separation between two adjacent particles.  Hence, the
half-way bounce-back algorithm with the coupling proposed by
Ladd~\cite{Ladd1994_1} and a first-order accurate time-integration scheme is
deemed most effective for applications involving large systems, where the
particles could become densely packed.

\subsection{Multi-component fluid interactions}\label{sec:method:shanchen}
To describe multi-component fluids we use the pseudo-potential method of Shan
and Chen~\cite{Shan1993,Shan1995}.  For each component $c$ we individually
solve Eq.~\ref{eq:method:boltzmann}.  In order
to couple the different components a repulsive fluid interaction force is added
between opposing fluid components $c$ and $c'$ at neighboring fluid nodes, as
given by
\begin{equation}
    \vec{F}^\textnormal{SC} = -\psi^c(\vec{x}_f, t) \sum\limits_{c'}G^{cc'}\sum\limits_i w_i \psi^{c'}(\vec{x}_f+\vec{c}_i, t) \vec{c}_i,
    \label{eq:method:shanchen}
\end{equation}
where $\psi^c$ is a pseudo-potential, and $G^{cc'}$ is the fluid
interaction strength.  For the current work we chose $\psi^c = 1-
\exp(-\rho^c/\rho_0)$ and limit the number of components to 2.

\begin{figure}
    \includegraphics{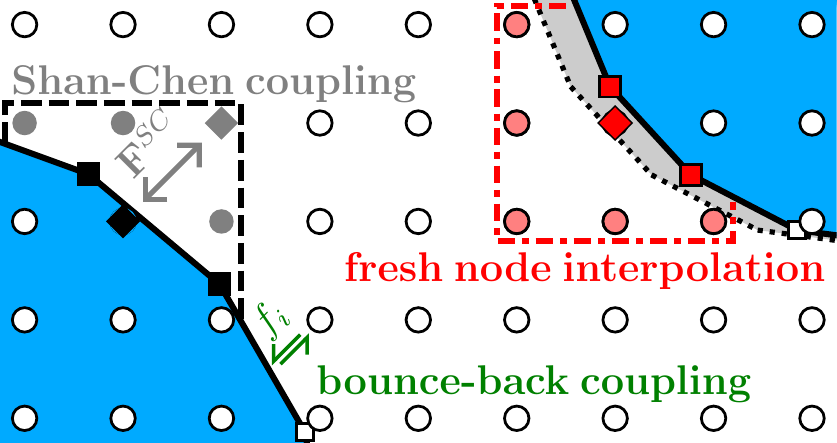}
    \caption{
(Color online) Schematic, two-dimensional visualization of the fluid-structure
coupling with the lattice of fluid nodes (circles+diamonds). The soft particles
(light blue areas) are represented as an infinitely thin membrane consisting of
$N_b$ boundary nodes (squares) and $N_f$ boundary elements (solid lines).  The
fluid encapsulated by the boundaries is decoupled from the bulk region by
bounce-back boundary conditions (dark green arrows).  In order to calculate fluid
interaction force $F^{SC}$ between fluid nodes on opposite sides of the
boundary (gray and black diamonds), we interpolate the fluid density on the
opposing side of the boundary. For the black diamond node this corresponds to
the gray nodes in the dashed box. The fluid interaction force that would be
applied at the interpolated node is distributed over the boundary nodes that
span the face over which the force would act (solid black squares).  When the
boundary moves, a fluid node can become uncovered by the particle (red
triangular node in gray shaded region). In such a case we interpolate the new
fluid density from the nearest neighboring nodes (red circles inside the
dotted box), where we ensure momentum conservation by adding the removed
momentum and subtracting the created momentum to the boundary nodes (solid red
squares) of the boundary element that crossed over the node.}
    \label{fig:method:overview}
\end{figure}

The interior and exterior fluid are decoupled at the boundary via the
bounce-back boundary conditions. Therefore, we also decouple the interior and
exterior fluid interaction forces.  In order to satisfy continuity close to the
boundary, we interpolate the neighboring fluid densities located on the
opposing side of a boundary
\begin{equation}
    \rho^c_\textnormal{bound}(\vec{x}_f,t) = \frac{1}{N^\textnormal{opp}_i}\sum\limits_i \rho^c(\vec{x}_f+\vec{c}_i,t),
    \label{eq:method:interpolateshanchen}
\end{equation}
where $i$ runs over all $N^\textnormal{opp}_i$ fluid nodes on the opposite side of the boundary as $\vec{x}_f$, as schematically indicated in Fig.~\ref{fig:method:overview}.
This way we interpolate the densities for the ring of fluid nodes just outside
our boundary for the inner fluid, and the ring of fluid nodes just inside the
boundary when calculating the fluid interaction force on the fluid nodes just
outside the boundary.  In order to conserve momentum the resulting force that
would act on the fluid node across the boundary is applied to the boundary
element that separates the set of fluid nodes.

This also allows to tune the contact angle of the particle surface by adding an
offset $\rho_\textnormal{virtual}^c$ to one component of the interpolated
densities $\rho^c_\textnormal{bound}$ of
Eq.~\ref{eq:method:interpolateshanchen}, as is proposed by Jansen et
al.~\cite{Jansen2011}.  Since the density offset is only added to one of the two
components, while the non-preferred fluid component is unchanged, we define the
dimensionless particle color $\Delta\rho$ by normalizing the added virtual
density offset $\rho_\textnormal{virtual}^c$ as
\begin{equation}
    \Delta\rho = \frac{1}{2}\frac{\rho_\textnormal{virtual}}{\rho_{\rm maj}-\rho_{\rm min}},
\end{equation}
where $\rho^c_{\rm maj}$ and $\rho^c_{\rm min}$ are the equilibrium majority and
minority densities corresponding to the chosen initial densities in the system
and the fluid interaction parameter $G^{cc'}$.
Here, a positive particle color indicates that the fluid density is added to the first fluid, whereas a
negative particle color indicates the density added to the other fluid.

The benefit of decoupling the interior and exterior regions by use of the bounce-back boundary conditions, is that we do not need a third fluid component for the interior region of the particles.
First, this will save a significant amount of computational resources.
Secondly, the interior region of the particle can be given the same composition as the exterior region, which makes it easier to initialize systems where the hydrostatic pressure inside the particle equals that outside the particle for an un-deformed mesh.
In the present work the interior of the particle is always initialized with the same fluid composition and fluid interaction strength as the exterior region when no fluid-fluid interfaces are present.
When fluid-fluid interfaces are present, the composition equals that of the lower fluid layer.
Careful checks validated that changing the interior composition to that of the upper fluid yields no difference in the presented results.

\subsection{Fresh fluid nodes}\label{sec:method:creation}\label{sec:methods:freshnodes}
One of the problems arising for moving boundaries is how to initialize fluid
nodes when a boundary element moves over it, commonly referred to as fresh node
treatment.  For solid particles one can choose not to remove the fluid that
becomes covered by a particle~\cite{Ladd1994_1}, however for fluid-filled
capsules this is not possible anymore.  In this case the local fluid at a fresh
fluid node $\vec{x}_\textnormal{fresh}$ needs to be removed and replaced by an
appropriate representation of the fluid present on the other side of the
boundary, whenever a boundary element passes a fluid node.

Several methods are possible for this, of which we use an interpolation
method~\cite{Aidun1998,Jansen2011}.  The new density at the fresh fluid node is
given by
\begin{equation}
    \rho^c_\textnormal{fresh}(\vec{x}_\textnormal{fresh},t) = \frac{1}{N^\textnormal{same}_i}\sum\limits_i \rho^c(\vec{x}_\textnormal{fresh}+\vec{c}_i,t),
    \label{eq:method:freshnodes}
\end{equation}
where the sum runs over all $N^\textnormal{same}_i$ neighboring fluid nodes
that are on the same side of the boundary as the fresh node, which are not
fresh nodes themselves during that time step.  This interpolation is
schematically visualized in Fig.~\ref{fig:method:overview}. As in Sec.~\ref{sec:method:bounceback}, the fresh fluid node is assigned the
average velocity of the three boundary nodes $\vec{x}_b$ of the nearest boundary element 
\begin{equation}
    \vec{v}(\vec{x}_\textnormal{fresh}, t) = \frac{1}{3} \sum\limits_{b=1}^{3} \vec{v}_b(t).
\end{equation}
In order to locally conserve the combined momentum of the fluids and particles,
the momentum of the removed fluid at the fresh node is added to the boundary
element that moved over the node, while the momentum of the newly generated
fluid is subtracted from the same element,
similar as to the momentum exchange method for solid objects as introduced by Ladd~\cite{Ladd2001}.

The interpolation as shown in Eq.~\ref{eq:method:freshnodes} however fails to
conserve the global mass in the system when there are forces or density
gradients close to the boundary, which is always the case when the particle is
near a fluid-fluid interface~\cite{Jansen2011,Frijters2012}.  Close to the
particle surface the fluid interaction force results in a slightly lower
density, and therefore the interpolation will always slightly over-estimate the
local density.  When the particle is close to the fluid interface, or has a
preferential contact angle other than 90$^{\circ}$, this effect increases
significantly.  Since no analytical solution is readily available for the
density profile around the surface, a correction which only uses local quantities is not feasible. Therefore, we apply an adaptive mass-correction term
which scales the density of fresh fluid nodes as~\cite{Jansen2011,Frijters2012}
\begin{equation}
    \rho^c_\textnormal{fresh} = \langle\rho^c\rangle \bigg(1 - C_0 \frac{\sum_c\rho^c_\textnormal{init}}{\rho^c_\textnormal{init}}\frac{\Delta\rho^c}{V_\textnormal{sys}} \bigg),
    \label{eq:method:masscorrection}
\end{equation}
where $V_\textnormal{sys}$ is the total volume of all fluid nodes in the
system, $C_0$ is a parameter used to tune the strength of the adaptive
correction, $\rho^c_\textnormal{init}$ is the average initial density of
component $c$ in the entire system, and $\langle\rho^c\rangle$ is the averaged
density obtained by Eq.~\ref{eq:method:freshnodes}.
We limit $\rho^c_\textnormal{fresh}$ between the smallest and largest density value of
its neighbors~\cite{Frijters2012}. 

\begin{figure}
    \includegraphics[width=0.9\linewidth]{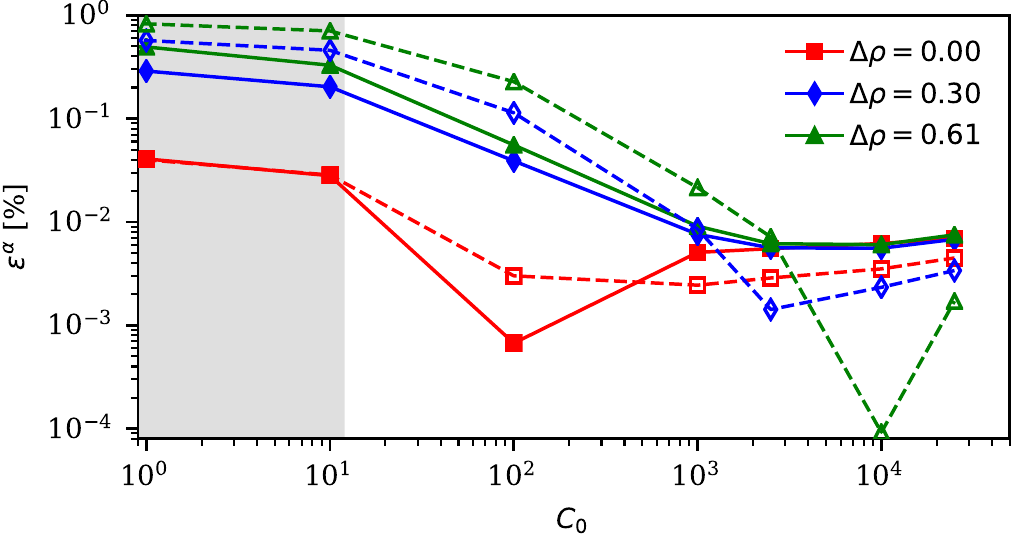}
    \caption{Relative errors in total mass for different wetting properties of the particle surface and different values of the mass-correction scheme of Eq.~\ref{eq:method:masscorrection}. A rigid spherical boundary with radius 10$\Delta x$ is initialised at a flat interface between two immiscible components, with a constant velocity of $v_x=10^{-3}$ parallel to the interface imposed on each individual boundary node, with a system size of (180, 60, 60) fluid nodes. The data is sampled after 250.000 simulation steps, i.e. on an integer number for the particle displacement to avoid capturing errors due to a change in discretisation of the particle on the fluid lattice. The open symbols indicate the relative error for the preferred component, while the solid symbols indicate the relative error for the non-preferred component. The lines are drawn as a guide to the eye. The region with the grey background indicates the simulations for which the error is still diverging after 250.000 simulation steps.
    }
    \label{fig:mass-conservation}
\end{figure}

In Fig.~\ref{fig:mass-conservation} we show the error in the total mass after 250.000 simulation steps for a system where a perfectly rigid particle adsorbed onto a fluid-fluid interface is propagated parallel to this interface. The particle movement is imposed by fixing the velocity of each boundary node to $10^{-3}$. Independent of the wetting properties of the particle, the relative error of each individual component is below 0.01\% when $C_0 \ge 2500$.
Hence, $C_0=2500$ is used throughout this work independent of the used particle colour.

\section{Results}\label{sec:results}
Throughout this entire section we fix the characteristic timescale $\tau$, the
reference density $\rho_0$, and the discretized time and length scales $\Delta
t$ and $\Delta x$ to unity. However, for clarity purposes we will denote the
non-dimensionalized lengths and times in units of $\Delta x$ and $\Delta t$.
Unless specified otherwise, the fluid interaction strength is set as
$G^{cc'}=3.6$, and the initial majority and minority densities are set as
$\rho^{\rm init}_{\rm maj}=0.7$ and $\rho^{\rm init}_{\rm min}=0.042$, which
yields a fluid-fluid surface tension of $\gamma^{cc'}=0.0388$.  The particles
are discretized with $N_f=2880$ boundary elements, and $N_b=1442$ boundary
nodes, where each boundary node is assigned a mass of 5$\rho_0 \Delta x^3$.
The fluid encapsulated by the particle is initialized with the same composition and fluid interaction strength as the lower fluid.

\subsection{Soft capsule in shear flow}
As a first validation of the fluid-structure coupling we simulate the
deformation of a capsule in a linear shear flow.
We define the capillary number $Ca$ as
\begin{equation}
    Ca = \frac{\rho^{tot}\nu \dot{\gamma} R_0}{\kappa_S},
\end{equation}
where $\rho^{tot}$ is the summed fluid density of both components, and $\dot{\gamma}$ is the shear rate.  We
initialize a single particle with a radius $R_0=7.64\Delta x$, a shear elasticity $\kappa_S$, and local extensibility $C=1$ (i.e. $\kappa_B=0$ and $B=0$) at the center of a cubic
domain of $48^3$ fluid nodes with a homogeneous binary fluid-mixture of
$\rho^c=\rho^{c'}=0.5$.  The interaction strength is set to $G^{cc'}$=0.0 mimicking a homogeneous mixture of non-interacting fluids.
A constant shear rate is imposed at the top and bottom $xy$-planes via
an on-site velocity boundary conditions~\cite{Zou1997,Hecht2010}.
The capillary number is varied by changing the shear elasticity while keeping the shear rate constant at $\dot{\gamma}=10^{-4}$.

\begin{figure}
    \includegraphics{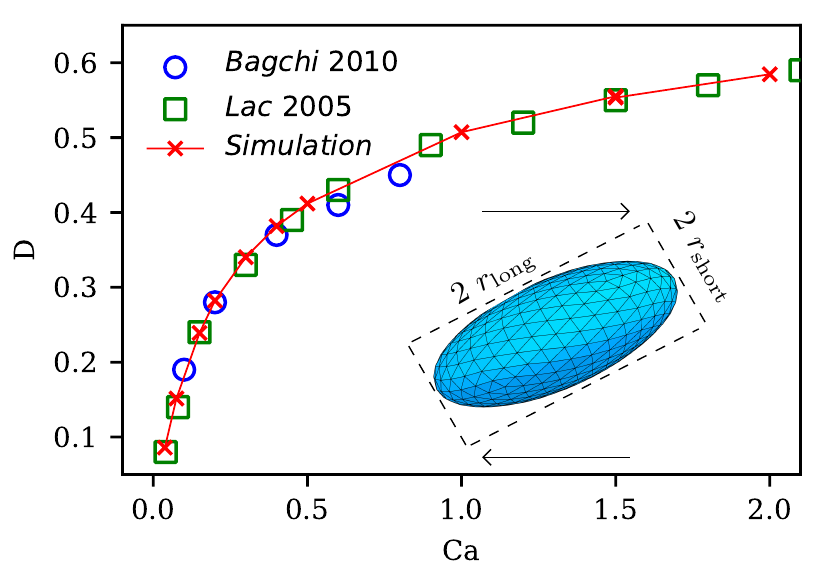}
    \caption{(Color online) Taylor deformation parameter $D$ for a soft capsule with local membrane in-extensibility $C=1$, homogeneous fluid densities $\rho^{c}=\rho^{c'}=0.5$, and interaction parameter $G^{cc'}=0.0$ for different capillary numbers $Ca$. 
    Our simulations show an excellent agreement with results reported previously in literature~\cite{Lac2005,Barthes2016,Bagchi2010}.
The inset visualizes the steady-state shape for $Ca=0.5$, where the dashed box indicates the short and long axis, and the arrows indicate the direction of the shear flow.}
    \label{fig:results:taylordeformation}
\end{figure}

For different shear rates we measure the steady-state Taylor deformation parameter
\begin{equation}
  D = \frac{r_\textnormal{long} - r_\textnormal{short}}{r_\textnormal{long} + r_\textnormal{short}},
  \label{eq:results:taylorparameter}
\end{equation}
where $r_\textnormal{long}$ and $r_\textnormal{short}$ are the length of the
long and short half-axis, respectively.  The lengths of these axes are obtained from
the tensor of inertia of the boundary surface \cite{Ramanujan1998}. As shown in
Fig.~\ref{fig:results:taylordeformation} our simulations show an excellent
agreement with the results reported by Lac et al.~\cite{Lac2005} and Bagchi et
al.~\cite{Bagchi2010} over a wide range of capillary numbers.  As expected,
this indicates that in the absence of a fluid interaction force we obtain the
correct physics when switching from a single-component fluid to a
multi-component fluid.

\subsection{Adsorption onto a fluid-fluid interface}\label{sec:interface:deformation}
In this section we show the capability of our method to model the adsorption of soft particles at fluid-fluid interfaces, and aim to characterize the error stemming from the coupling between the multi-component fluid and the bounce-back boundary conditions for soft particles.
Soft particles adsorb onto nearby fluid-fluid interfaces in order to minimize the interfacial energy.
For rigid particles the interplay between the inertia of the particle and the surface tension of the fluid-fluid interface dominates the adsorption dynamics.
Soft particles have another degree of freedom, namely the softness of the particle.

All simulations in this section are performed in a system with a fluid-fluid
interface in the $xy$-plane, located at $z_\textnormal{interface}=32\Delta x$
in a cubic domain of $(64 \Delta x)^3$ fluid nodes. The system is closed at the
top and bottom with rigid boundaries using the half-way bounce-back scheme, and
periodic boundary conditions in the other directions.

We define the dimensionless softness parameter of the capsule at a fluid-fluid
interface as $\beta=R_0^2\gamma^{cc'}/\kappa_B$.  Unless specified otherwise we
initialize the particles with an initially spherical shape with $R_0=10\Delta
x$ and $N_f=2880$ boundary elements.  The volume conservation modulus is chosen
on purpose at a large value of $\kappa_V=20$ to ensure a quasi
volume-incompressible capsule with fluctuations in the encapsulated volume in
the order of 0.01\% for all presented simulations. The exact choice here is
however not significant for the presented work, and smaller moduli are possible
as well.

\begin{figure}
    \includegraphics[width=0.45\textwidth]{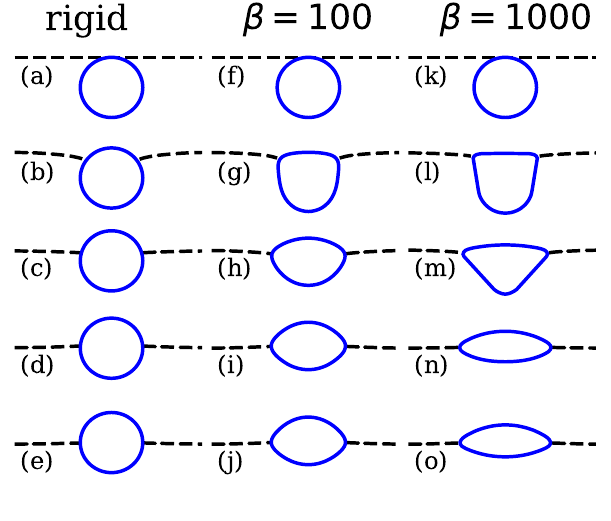}
    \caption{(Color online) Visualization of adsorption of a particle (blue solid line) onto a fluid-fluid interface (black dashed line), on a slice perpendicular to the fluid-fluid interface centered at the particle center. The system is initialized with the top of the particle touching an undisturbed fluid-fluid interface. The top panels show the initial state at $t=0$, and each consecutive frame below shows the systems 400$\Delta t$ later. (a-e) Adsorption of a perfectly rigid particle. (f-j) Adsorption of a particle with softness parameter $\beta=100$. (k-o) Adsorption of particle with $\beta=1000$. Due to the inertia of the particle boundary, the fluid-fluid interface can be observed to be pulled downwards in the second and third rows of panels, whereas it is slightly pulled upwards in the fourth and fifth rows.}
    \label{fig:interface:adsorption}
\end{figure}

In Fig.~\ref{fig:interface:adsorption} we visualize the adsorption of a soft particle with $\beta=100$ and $\beta=1000$ onto a fluid-fluid interface, compared against a rigid particle.
For now we focus qualitatively on the dynamics of an adsorbing particle, and therefore only visualize the first 1600$\Delta t$ time steps, in which the particles adsorbs onto the fluid-fluid interface and still shows a small, dampened movement around its final equilibrium position.

As expected from a physical viewpoint, we observe for all three cases a simultaneous downwards motion of the interface towards the particle surface, while the top of the particle extends at the fluid-fluid interface under the outward pull of the surface tension.
Increasing the softness parameter appears to mitigate the downward movement of the interface, where the height difference of the fluid-fluid interface at the contact point and at the point farthest away from the particle is 2.3$\Delta x$ for the rigid particle, 1.8$\Delta x$ and $1.1\Delta x$ respectively for the particle with $\beta=100$ and $\beta=1000$.
This can be qualitatively explained by the fact that for very soft particles it is energetically more favorable to match the contact angle between the particle surface and fluid-fluid interface via deformation of the surface than via a deformation of the interface.
It has to be noticed that the system-size here is relatively small to study such deformation effects.
The capillary length can be multiple times larger than the particle radius, which requires larger system sizes.
A more thorough and quantitative study of such effects is left for future work, and beyond the scope of the current paper.

Many different studies have already been performed on the accuracy of various constitutive models for the dynamics of soft particles.
For instance, Ramanujan studied the deformation of capsules for various fluid viscosities~\cite{Ramanujan1998}, Omori studied the correlation between spring network models and continuum constitutive laws~\cite{Omori2011}, Barth{\`e}s-Biesel extensively studied the deformation of various types of capsules~\cite{Barthes2016}, and the discretization of bending algorithms was studied by both Guckenberger~\cite{Guckenberger2017} and Tsubota~\cite{Tsubota2014}.
Hence, we purposefully attempt to isolate the error due to the fluid-structure coupling, which is the new part proposed in this work, from those stemming from the method used to model the elastic boundaries.

\begin{figure}
    \includegraphics{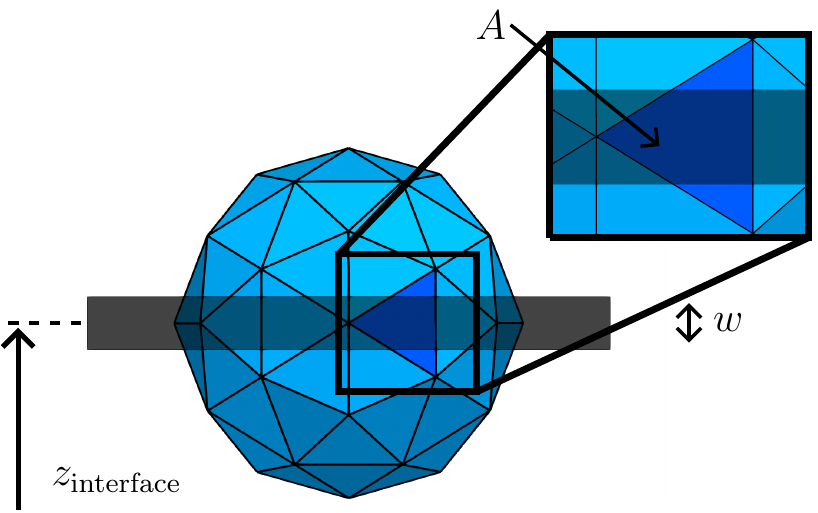}
    \caption{(Color online) Schematic visualization for determining the emulated tensile force from the fluid-fluid surface tension acting on a boundary element which is sliced by the interface position in the used reference model. The interface is assumed to have a width $w$ around the interface position $z_\textnormal{interface}$ (gray box). We divide the desired surface tension based on the overlapping area $A$ of the boundary element (dark blue) and interface region, and divide the force evenly over the three boundary nodes that span the triangular boundary element.}
    \label{fig:interface:referencemodel}
\end{figure}

For this purpose we design a reference model that uses the same routines as our
implemented code as presented in section~\ref{sec:method}, but disables the
hydrodynamics by ignoring all routines related to the LBM.  Instead of the
hydrodynamic force of LBM, we directly apply the tensile force
$F^\textnormal{tensile}$ of the surface tension acting on the particle surface
on all boundary elements within a width $w$ of a stationary emulated position
of the interface.  The emulated tensile force on a boundary element is computed
from the area of the boundary element covered by the emulated interface as
\begin{equation}
    F^\textnormal{tensile} = \frac{\gamma A}{w},
  \label{eq:interface:emulatetension}
\end{equation}
where $A$ is the area of the boundary element which lies between two planes
within a distance $w/2$ of the emulated interface position
$z_\textnormal{interface}$, and $\gamma$ is the fluid-fluid surface tension.
The determination of the area $A$ is schematically
visualized in Fig.~\ref{fig:interface:referencemodel}.  The emulated tensile
force is distributed homogeneously over the three boundary nodes which span the
boundary element.  We set the width $w=1\Delta x$, equal to the
characteristic length scale of the fluid lattice, unless specified otherwise.

The reference model uses exactly the same elastic laws, the same
discretization of the particle surface, and the boundary nodes are updated with
the same time integration scheme.  The input values for the surface tension are
obtained from separate simulations mimicking a Laplace test, which relate the
pressure difference of a droplet for a specific interaction parameter $G^{cc'}$
with the droplet radius (data not shown)~\cite{Frijters2012}. This reference model a priori
excludes sources of errors other than the fluid-structure coupling, and allows
us to compare the simulation results directly against the deformation of any
constitutive elastic law already established in the literature.

\begin{figure}
    \centering
    \includegraphics{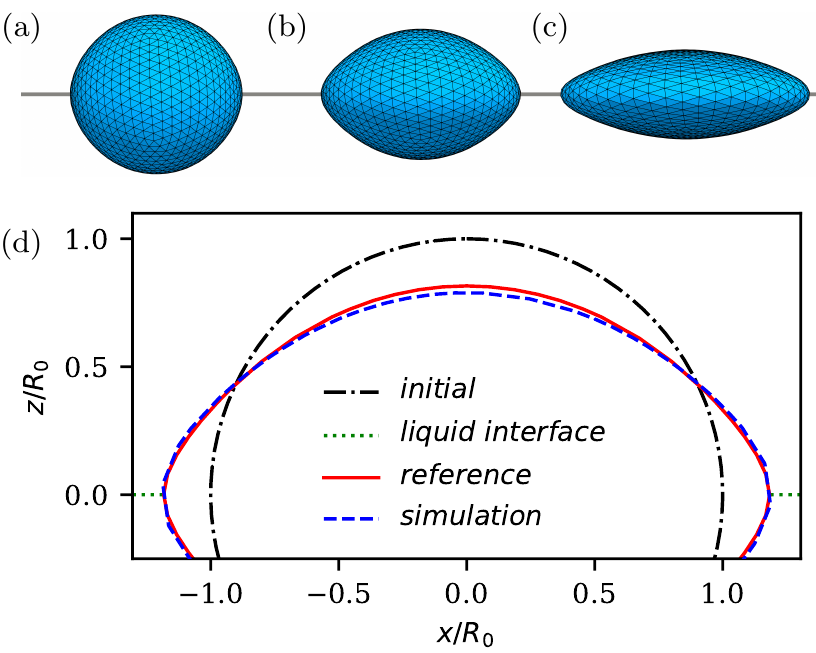}
    \caption{(Color online) Equilibrium shapes of a soft capsule adsorbed onto a fluid-fluid interface. (a-c) The equilibrium shape for a capsule with $\nu_s=0.9$ and softness parameter of respectively $\beta=[10, 100, 1000]$. The fluid-fluid interface (not shown) runs horizontal through the center of the particles. (d) Overlap of equilibrium shape of our numerical method (dash-dotted line) and the shape obtained with the reference model of Eq.~\ref{eq:interface:emulatetension} (solid line) for $\beta=100$ and $\nu_s=0.9$. The dotted line shows the initial, stress-free shape of the capsule, and the green shaded area indicates the lower fluid component.}
    \label{fig:interface:shapes}
\end{figure}

To quantify the error of our results we use the extrema of the particle at the
poles $r_p$ and at the fluid-fluid interface $r_i$, where $r_p$ is the
equilibrium half-axis in the plane perpendicular to the fluid-fluid interface, and
$r_i$ the equilibrium half-axis in the plane parallel to the fluid-fluid
interface. The errors are then defined as the normalized difference between
the reference case and our simulation method as
\begin{equation}
\begin{aligned}
    \Delta_p = \frac{\langle r^\textnormal{ref}_{p} \rangle - \langle r^\textnormal{sim}_{p} \rangle}{R_{0}}, \\
    \Delta_i = \frac{\langle r^\textnormal{ref}_{i} \rangle - \langle r^\textnormal{sim}_{i} \rangle}{R_{0}},
\end{aligned}
    \label{eq:interface:error}
\end{equation}
since due to the bending rigidity the entire shape is characterized by these
extrema in equilibrium.  The extrema $r_p$ and $r_i$ can be obtained directly from
the tensor of inertia, similar as done for Taylor deformation parameter $D$ of
Fig.~\ref{fig:results:taylordeformation}. In this manner the errors $\Delta_p$
and $\Delta_i$ provide an intuitive dimensionless measure for the error
stemming from the fluid-structure coupling.

In Fig.~\ref{fig:interface:shapes}, we visualize the obtained equilibrium shapes for a Poisson ratio $\nu_s=0.9$ and softness parameters $\beta$ of 10, 100, and 1000 to elucidate the used parameters.
Furthermore, we visualize a comparison of the obtained equilibrium shape
between our simulation method and the reference simulation for $\beta=1000$.  The particle is stretched to
$r^\textnormal{sim}_i=11.55\Delta x$ and $r^\textnormal{sim}_p=7.49\Delta x$,
with corresponding errors of $\Delta_p=-2.2\cdot10^{-2}$ and
$\Delta_i=1.6\cdot10^{-2}$.

\begin{figure}
    \includegraphics{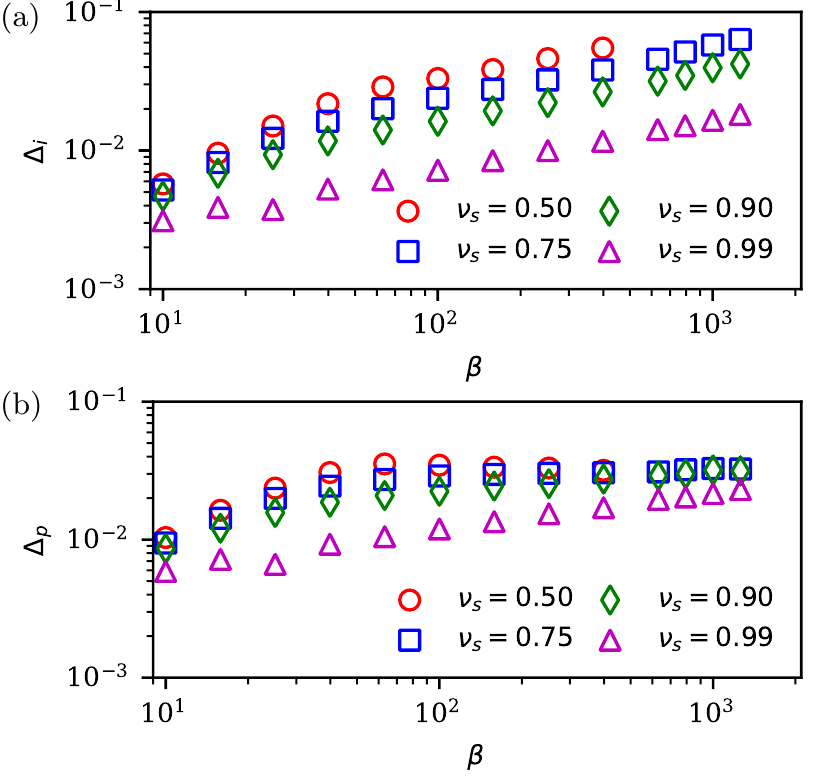}
    \caption{(Color online) Characterization of the error at (a) the interface $\Delta_i$ and (b) poles $\Delta_p$ in the equilibrium shapes as defined in Eq.~\ref{eq:interface:error} for different softness parameters $\beta$ and Poisson ratio $\nu_s$ for a particle with radius $R_0=10\Delta x$.
    }
    \label{fig:interface:overlap-beta}
\end{figure}

Over a wide range of softness parameters $\beta$ and Poisson ratios $\nu_s$
we observe an excellent agreement with the reference model, as shown in
Fig.~\ref{fig:interface:overlap-beta}.  For $\beta=1000$ and $\nu_s=0.9$ the ratio between the axis in
the plane perpendicular and parallel to the fluid-fluid interface is 2.3, with an absolute mismatch at the fluid-fluid interface of only $\langle
r^\textnormal{ref}_{i} \rangle - \langle r^\textnormal{sim}_{i} \rangle =
0.31\Delta x$, which is considered as excellent agreement when considering the
accuracy of the used half-way bounce-back boundary conditions.

\begin{figure}
    \includegraphics{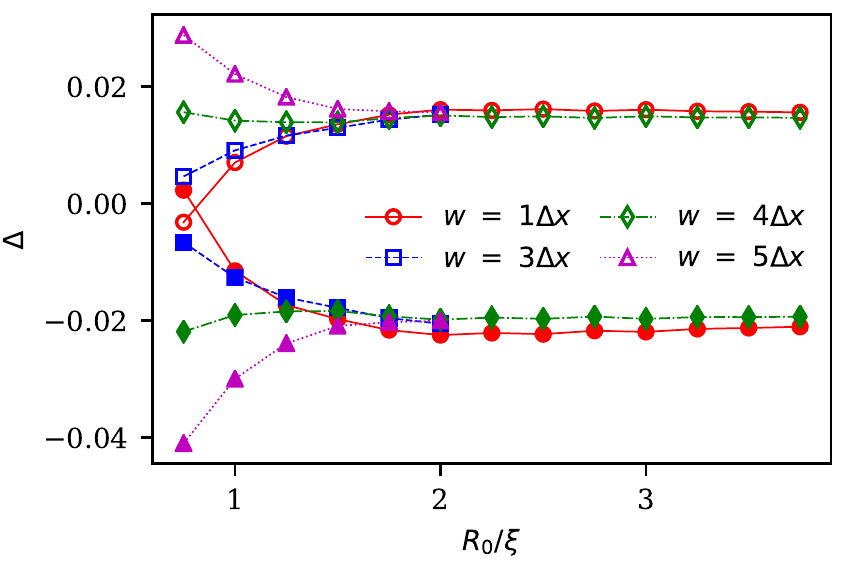}
    \caption{(Color online) Characterization of the errors in the equilibrium shapes as
        defined in Eq.~\ref{eq:interface:error}, for a capsule with fixed softness
        parameter $\beta=100$ and Poisson ratio $\nu_s=0.9$ for different initial radii
        $R_0$ and emulated interface width $w$. The open symbols indicate the error at
        the fluid-fluid interface $\Delta_i$, and the solid symbols the error at the
        poles $\Delta_p$. For small radii we clearly observe the influence of the
        diffuse interface stemming from the pseud-potential LBM, whereas we achieve a
        roughly constant error for different radii when using an interface width of
        $w=4\Delta x$ for the reference model. The lines for $w=3\Delta x$ and $w=5\Delta x$ are
        cropped after $R_0/\xi=2$ for clarity, since they closely follow the line for
        $w=4\Delta x$ in this regime.}
    \label{fig:interface:overlap-radius}
\end{figure}

The increasing error for larger softness parameters can be understood by
considering the minor-axis of the particle perpendicular to the fluid-fluid
interface, which decreases for an increased softness and approaches the diffuse
interface width $\xi$ of the multi-component model for very large softness
parameters.
The largest deformation is observed for a particle with $\beta=1259$ and
$\nu_s=0.75$, where the minor axis decreases down to 3.6$\Delta x$ and the
errors with an emulated width of 1 read $\Delta_p=-3.3\cdot10^{-2}$ and
$\Delta_p=6.3\cdot10^{-2}$ at the poles and interface respectively.  The
typical width of the diffuse interface for the used interaction parameter
$G^{cc'}=3.6$ is $\xi\simeq4\Delta x$.  When we increase the emulated width $w$
of the interface, we indeed observe a decrease in the error at both the poles
and the interface, where for an emulated width of $w=4$ the errors reduce to
$\Delta_p=-2.2\cdot10^{-2}$ and $\Delta_p=3.3\cdot10^{-2}$.  As can be seen in
Fig.~\ref{fig:interface:overlap-radius} the errors for particles with a very
large initial radius are only weakly dependent on the particle radius and
appear to approach a uniform value independent of the emulated interface width
when $R_0 \gg \xi$.  For smaller intermediate particle radii $R_0 \lesssim \xi$
the importance of the diffuse interface becomes apparent, where the errors with
a very low emulated width $w$ become strongly dependent on the ratio of the
particle radius and diffuse interface width.

\begin{figure}
    \includegraphics{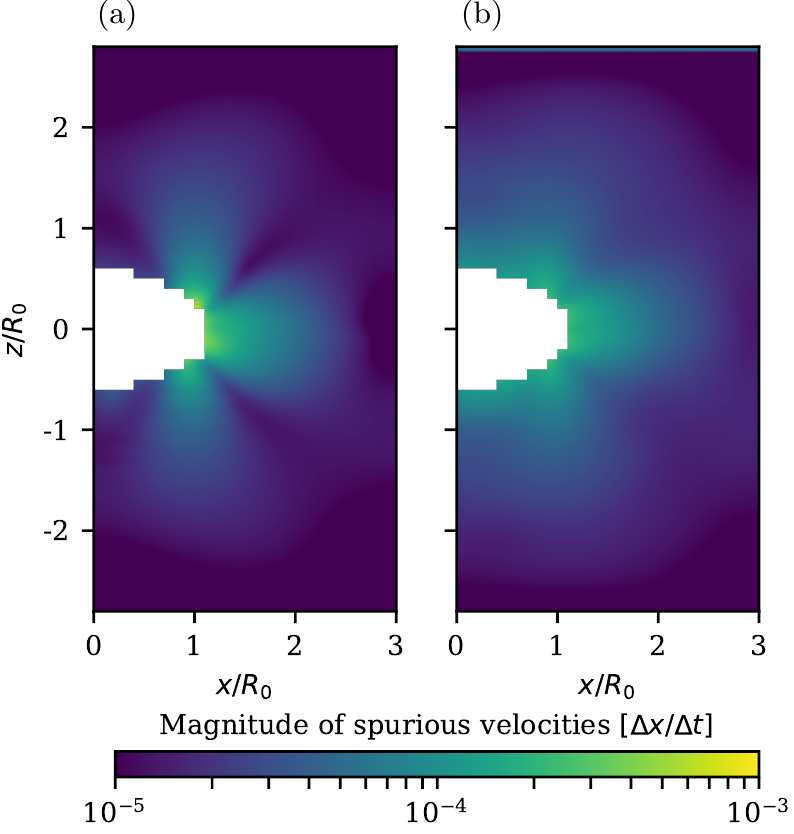}
    \caption{(Color online) Spurious velocities near the contact point of a soft particle with a fluid-fluid interface.
            Due to the curved boundary the pseudo-potential multi-component fluid method will always yield spurious velocities near the contact point with the fluid interface.
            The velocity field outside the particle boundary is visualized on a vertical slice through the center of the particle, where the fluid nodes encapsulated by the boundary are shown in white.
        (a) Magnitude of the spurious velocities around a stationary soft particle with $\beta=500$ and $\nu_s=0.9$.
        (b) Magnitude of the spurious velocities around a particle co-moving with the reference frame at a velocity of $\vec{v}=(0.001, 0.002, 0.0)$. The reference frame is driven with a constant velocity $\vec{v}$ parallel to the fluid interface by using on-site velocity boundary conditions on the top and bottom planes.
        As a result of the changing discretisation of the particle boundary on the fluid lattice, the averaged spurious velocities are smeared out.
    }
    \label{fig:spurious-currents}
\end{figure}

To verify the Galilean invariance of the presented method, a soft particle with $\beta=500$ and $\nu_s=0.9$ adsorped onto a fluid-fluid interface is studied where the top and bottom plane of the system are driven with a constant velocity $\vec{v}=[0.001, 0.002, 0]\Delta x/\Delta t$ parallel to the interface.
In order to exclude effects from the initial acceleration of the particle, the simulation is run for 500,000 simulation steps, of which the last 50,000 are sampled every 5,000 steps.
No observable differences were found in the particle shape, and the pressure inside and around the particle only shows a relative difference of 0.03\%.
Due to the pseudo-potential multi-component fluid method, spurious currents are present near the contact point of the particle boundary with the fluid-fluid interface, similar to the spurious currents that are known to persist indefinitely near the curved fluid-fluid interfaces.
Small differences occur in the fluid velocities around the contact point on the time scale of $\Delta t$ as a result of the continuously changing discretisation of the particle boundary on the fluid lattice.
These stem from the half-way bounce-back scheme, which is known to induce local force fluctuations whenever a boundary element jumps over a fluid node.
In Fig.~\ref{fig:spurious-currents} we compare the magnitude of the spurious velocities for a stationary particle and a particle moving along with the system with velocity $\vec{v}$.
Here, it can be observed that the spurious currents do not show an increase in magnitude, but do show a minor increase in their extent.
However, for the current work we deem the accuracy of the half-way bounce-back scheme sufficient.
In future works, when a higher accuracy is required, the errors stemming from the changing discretisation of the particle boundary on the fluid lattice could be mitigated by using a fluid-structure coupling scheme with a higher accuracy.
Similarly, the effect of the spurious currents could be reduced by exchanging the multi-component model for a scheme with reduced spurious currents.

\subsection{Soft capsule at an interface under shear}
In this section we briefly demonstrate how the presented method can be used to study the dynamics of a system with shearing boundary conditions with a soft capsule adsorbed onto a fluid-fluid in the center of the system, where the shear is applied parallel to the interface.
    We initialize the system domain with 180 fluid nodes in both directions parallel to the fluid-fluid interface and 87 fluid sites perpendicular, where the odd numbered system height is chosen in order to easily initialize the interface exactly half-way between the shearing boundaries.
    To disable fluid-fluid interactions between the top and bottom plane these are initialized with non-moving bounce-back boundaries.
    On-site velocity boundary conditions are applied on the two planes of fluid nodes next to these boundaries to impose the shear velocities~\cite{Zou1997,Hecht2010}.

\begin{figure}
    \includegraphics{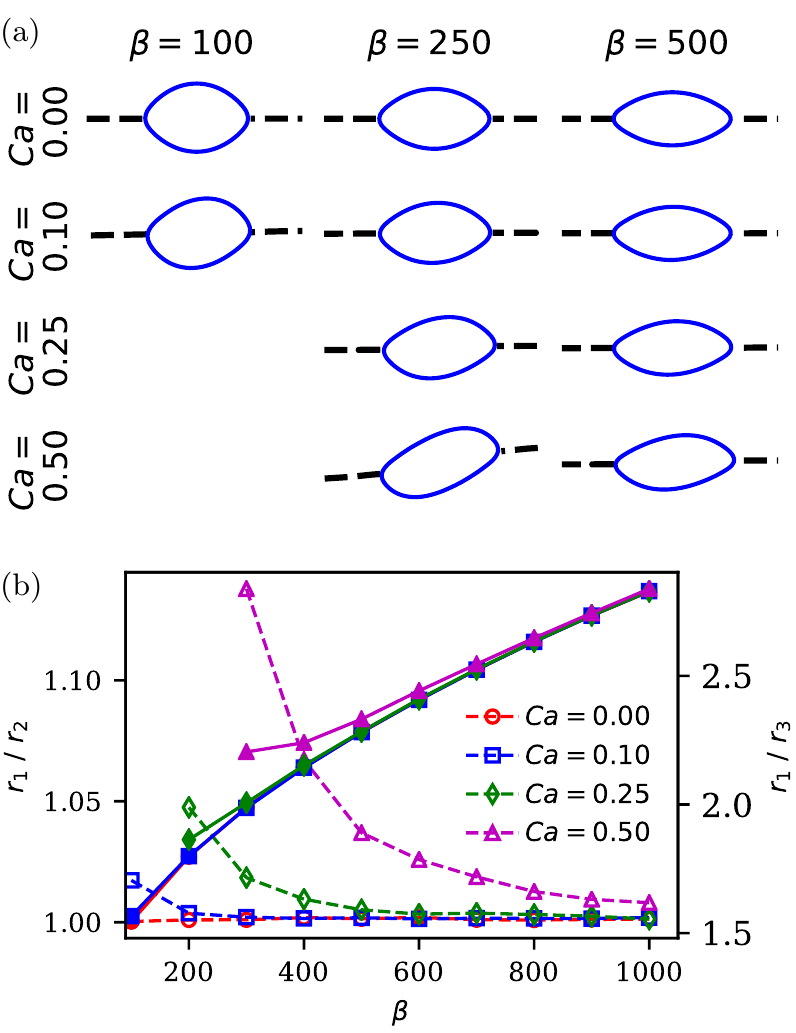}
    \caption{(Color online) Deformation of a soft particle adsorbed onto a fluid-fluid interface with a shear velocity applied parallel to the interface.
            (a) Steady-state particle shapes (blue solid line) and fluid-fluid interface (black dashed line) for different capillary numbers and softness parameters, visualized as a 2D slice through the center of the particle. The fluid flows to the right at the top of the system, and to the left at the bottom of the system. For increasing capillary numbers the asymmetry in the particle shape increases, where for the lower softness parameters the interface starts to deform as a result of the dynamical wetting of the tank-treading particle boundary.
            (b) Ratios between the semi-axes $r_1$ and $r_2$ (dashes lines and open markers) and $r_1$ and $r_3$ (solid lines and symbols) for different capillary numbers and softness parameters. The semi-axes are labeled in descending order labeled as $r_n$.
            For increasing softness parameters it can be observed that the influence of the shear flow reduces, and the ratio between $r_1$ and $r_3$ converges to the symmetric case of $Ca=0.0$.
    The last shapes for $\beta=100$ and $Ca=[0.25, 0.5]$ are not included since the needed shear velocities approach the numerical limits of the LBM.
    }
    \label{fig:interface:shear-interface}
\end{figure}

When no shear is applied, i.e. $Ca=0$, the particle has two symmetric major semi-axes parallel to the fluid-fluid interface and one minor semi-axis perpendicular to the interface, similar as previously visualized in Fig.~\ref{fig:interface:shapes}.
In the top panel of Fig.~\ref{fig:interface:shear-interface} we show the steady-state shape of the particle and the interface for different capillary numbers and softness parameters.
Here, it can be observed that the symmetry is broken when the system is sheared, and the major semi-axis parallel to the shear flow elongates.
In the bottom panel of Fig.~\ref{fig:interface:shear-interface} we visualize the ratio between these two axes (the solid lines and symbols), and it can be clearly seen that the asymmetry for the two largest axes increases for increasing capillary numbers and decreasing particle softness.
Simultaneously, the ratio between the major and minor axis deviates stronger from the base curve set by $Ca=0.0$ for lower softness parameters when the capillary number is increased, while for large softness parameters the ratio becomes independent of the capillary number.
Intuitively this can be explained by the fact that softer particles have less area perpendicular to the shear flow, and therefore experience a lower stress from the shear flow.
This results in a lower tank-treading velocity of the membrane, which is indeed observed in the simulations (data not shown).

\begin{figure}
    \includegraphics[width=0.45\textwidth]{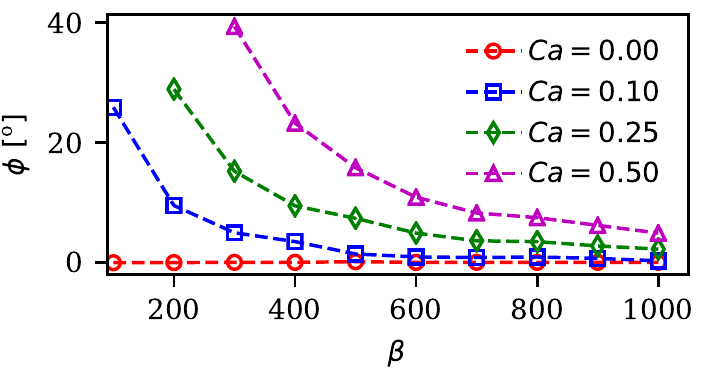}
    \caption{(Color online) Angle of largest semi-axis with the plane of the undisturbed interface for a soft particle adsorbed onto a fluid-fluid interface with a shear velocity applied parallel to the interface for different capillary numbers and softness parameters. For increasing capillary numbers the symmetry of the particle is broken as shown in Fig.~\ref{fig:interface:shear-interface}, and the angle between the major axes and the horizontal plane increases.}
    \label{fig:interface:shearing-angle}
\end{figure}

In Fig.~\ref{fig:interface:shearing-angle} we show the angle $\phi$ between the largest semi-axis and the horizontal plane (i.e. the undisturbed fluid-fluid interface), which increases for increasing capillary numbers and decreasing softness parameters.
This rotation of the inertial tensor semi-axes and the corresponding asymmetry in the particle shape also influences the fluid-fluid interface, as can be seen in Fig.~\ref{fig:interface:shear-interface} for $Ca=0.5$ and $\beta=250$.
Here one can observe the dynamical wetting of the particle surface, where the interface rises on one side of the particle while lowering on the other side in a symmetric manner due to the rotational velocity of the particle surface.
Qualitatively this fits with the expectations from experiments and simulations on similar systems with rigid particles~\cite{Capelli2015}, where the additional asymmetric deformation of the particle appears to allow a stronger deformation of the interface.
A more detailed and thorough analysis of this phenomenon, including the effect of the preferential contact angle of the particle boundary, is left for future work.

\subsection{Tunable contact angles}\label{sec:interface:contactangle}
Similar to the work of Jansen et al.~\cite{Jansen2011} our model is capable
of tuning the preferential contact angle $\theta^{eq}$ of the particle surface
with the fluid-fluid interface. A positive particle color $\Delta\rho$ is
expected to result in the particle preferring one phase over the other phase,
and a change from the $\theta^{eq}=90^\circ$ contact angle. For a rigid sphere
with radius $R_0$ the contact angle can be measured by the distance $\delta z$ between the
center of mass of the particle and the position of the interface as
\begin{equation}
    \cos\theta = \frac{\delta z}{R_0}.
    \label{eq:interface:contactangle}
\end{equation}

\begin{figure}
    \includegraphics{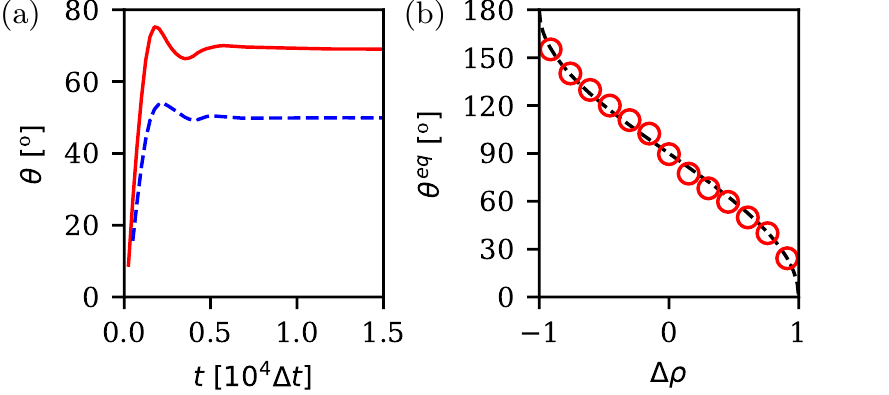}
    \caption{(Color online) Measured contact angle $\theta$ in degrees of rigid spherical boundaries with radius $R_0=10\Delta x$.
    (a) Evolution of instantaneous contact angle as measured with
    Eq.~\ref{eq:interface:contactangle} towards the equilibrium angle for a
    rigid capsule with a particle color of $\Delta\rho$=0.30 and 0.61. To
    demonstrate the effect of the particle inertia the particles are initialized
    inside their preferred fluid with the top of the particle touching the
    fluid-fluid interface.
    (b) Equilibrium contact angle as measured with
    Eq.~\ref{eq:interface:contactangle} for different particle colors
    $\Delta\rho$ for rigid capsules. The dashed line gives the line $\theta=180/\pi
    \cdot \cos^{-1}(\Delta\rho)$, which shows good agreement with the obtained
    equilibrium contact angles from the simulation (red circles).
    }
    \label{fig:interface:contactangle}
\end{figure}

The particle is initially submerged in its preferred fluid component
with the particle surface penetrating the fluid-fluid interface by roughly 1$\Delta x$.  A rapid movement
towards the equilibrium position, with some minor oscillations as a result of
the inertia of the particle, can be observed as shown in
Fig.~\ref{fig:interface:contactangle}.

From Young's equation the contact angle is given by the balance of the interfacial tensions $\gamma_{pc}$ and $\gamma_{pc'}$ between the particle and two fluid components, and the tension $\gamma_{cc'}$ between the two fluid components
\begin{equation}
    \cos\theta = \frac{\gamma_{pc} - \gamma_{pc'}}{\gamma_{cc'}}.\label{eq:results:young}
\end{equation}
For a neutral particle color $\Delta\rho=0$ the fluid-structure tensions are
symmetric and negate the nominator of Eq.~\ref{eq:results:young}.  The
virtual fluid density added to the fluid interaction force by a non-zero
particle color increases the asymmetry in the two fluid-structure interfacial
tensions and can be expected to yield a scaling of the equilibrium contact
angle as
\begin{equation}
    \theta \propto \cos^{-1}(\Delta\rho).
    \label{eq:results:anglescaling}
\end{equation}
As seen in the right panel of Fig.~\ref{fig:interface:contactangle}, the
measured equilibrium contact angle $\theta^{eq}$ of a rigid spherical boundary
shows a good agreement with Eq.~\ref{eq:results:anglescaling}.

\begin{figure}
    \includegraphics{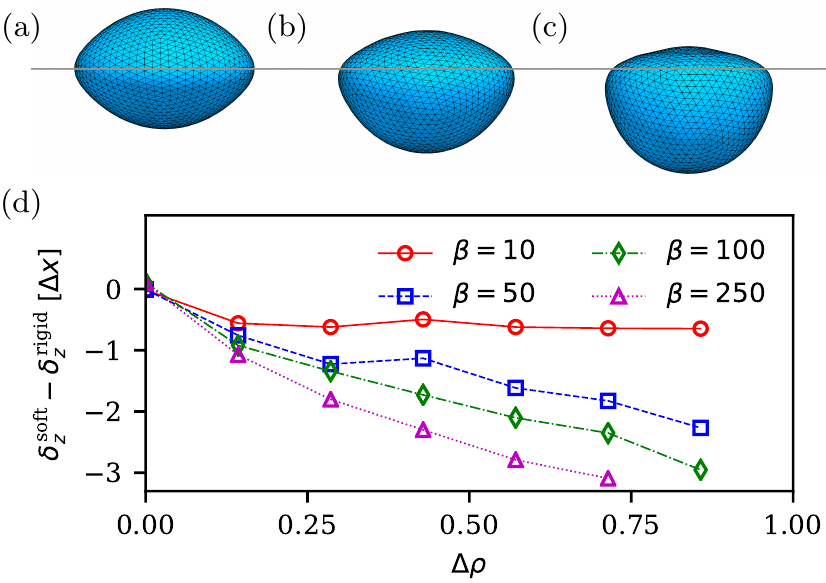}
    \caption{(Color online) Vertical displacement $\delta_z$ of the particle center of mass relative to the interface for increasing particle color $\Delta\rho$.
        (a-c) Equilibrium shapes of a soft capsule adsorbed onto an interface with $B=0.01, \beta=100, \nu_s=0.9$, and a particle color of respectively $\Delta\rho=[0.0, 0.30, 0.61]$ (from left to right).
        The fluid interface is shown by the grey line.
        (d) The difference in displacement of a soft particle $\delta^\textnormal{soft}_z$ and a rigid particle $\delta^\textnormal{rigid}_z$ for different softness parameters $\beta$.
        Soft particles are observed to penetrate less far into the preferred fluid as compared to rigid particles with the same preferred contact angle.
        For a very low softness $\beta=10$ the difference in displacement reaches a constant value particle colors $\Delta\rho \ge 0.15$.}
    \label{fig:interface:soft_contactangle}
\end{figure}

For rigid particles at a fluid-fluid interface the vertical displacement
$\delta_z=|z_\textnormal{interface}-z_\textnormal{particle}|$ is the only
mechanism that allows for the match of the preferential contact angle
$\theta^{eq}$ of the particle surface without deforming the interface. Soft
particles, however, can also deform in order to match the preferential
contact angle. Thus, for larger values for the particle color the
equilibrium shape turns into a hazelnut shape, as depicted in
Fig.~\ref{fig:interface:soft_contactangle}.  This deformation simultaneously
decreases the vertical displacement $\delta_z$ needed to match the preferential
contact angle with the interface. In
Fig.~\ref{fig:interface:soft_contactangle} we clearly observe the influence
of the particle softness $\beta$ on the displacement as compared to a rigid
particle with an equal particle color $\Delta\rho$, where the particle
displacement $\delta_z$ reduces for increasing softness parameters. For
 $\beta=100$ and $\Delta\rho=0.91$ the
difference in displacement is almost 30\% of the particle radius, while
for relatively stiff capsules with $\beta=10$ the change in vertical
displacement is only observed for low particle color values and reaches a
constant difference for $\Delta\rho > 0.15$.

\subsection{A droplet covered with soft particles}\label{sec:interface:outlook}
To further demonstrate the capability of our method to simulate the interplay between soft particles and fluid-fluid interfaces, 
we simulate
the covering of a droplet of radius $17.5\Delta x$ suspended in another fluid by 24 soft
capsules with $\beta=100$, $\nu_s=0.9$, and $R_0=7\Delta x$.
The particle centers are initialized on a sphere with
radius $R=23\Delta x$, such that each particle slightly penetrates the droplet.
Two different simulations are shown, one with hydrophilic
particles with $\Delta\rho=-0.61$ and another with hydrophobic particles with
$\Delta\rho=0.61$.  A short range repulsive hard-core particle-particle
interaction force is added between boundary nodes of opposing particles to
prevent the overlap of the particles
\begin{equation}
    F^\textnormal{P-P}(d) = \begin{cases}
        \epsilon^\textnormal{P-P} (d_0 - d)^{3/2} & d < d_0, \\
        0 & d \geq d_0,
    \end{cases}
\end{equation}
where $\epsilon^\textnormal{P-P}$ is a constant parameter fixed here to $0.5$, $d$
is the separation distance between the two boundary nodes, and $d_0 = 1.25\Delta x$ is a cut-off
distance for the interaction force.  The exact shape of the repulsive force
is however unimportant as long as it is sufficiently strong to prevent the
overlap of the particles.
\begin{figure}
    \includegraphics{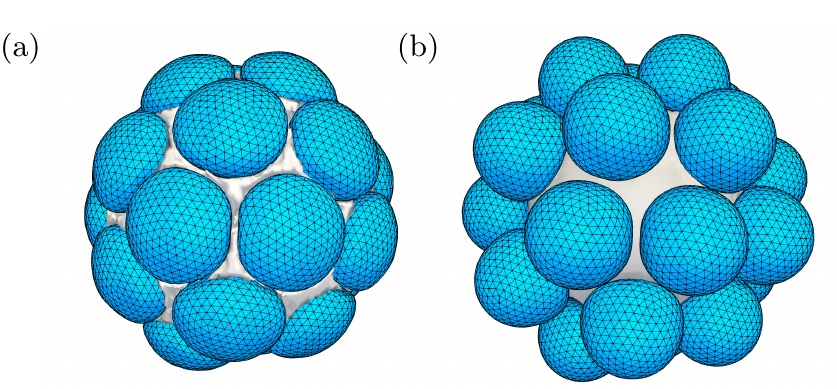}
    \caption{(Color online) Comparison of the equilibrium shapes of adsorbed soft particles at
a droplet for particles with a particle color of (a) $\Delta\rho=0.61$ and (b)
$\Delta\rho=-0.61$. The soft particles (blue) are initialized in a spherical
configuration with $R=23\Delta x$, centered around the droplet. All particles
have an initial radius $R_0=7\Delta x$, softness parameter $\beta=100$, and
Poisson ratio $\nu_s=0.9$. The droplet with an initial radius of $R=17.5\Delta
x$ is visualized with the iso-density line $\rho^\textnormal{drop}=0.5$ (light
gray), while the suspending bulk fluid is not visualized.}
    \label{fig:results:showcase}
\end{figure}

In Fig.~\ref{fig:results:showcase} we compare the obtained configurations of
both simulations after $5\cdot10^3 \Delta t$. The hydrophilic particles are
almost completely pulled inside the droplet, covering nearly the entire surface
of the droplet.  Due to the close-packing at the interface, the hydrophilic
particles cannot obtain the preferred hazelnut shape as shown in
Fig.~\ref{fig:interface:soft_contactangle}, and the deformation is also
significantly influenced by the interactions with other particles.  The
hydrophobic particles only slightly penetrate the surface of the droplet,
resulting in a more compact droplet and a larger surface area uncovered by
particles.  A detailed investigation of this system goes beyond the scope of
this work and will be presented in a future paper.

\section{Conclusion and discussion}\label{sec:conclusion}
We have presented a method capable of simulating soft particles in
multi-component fluids, where we combined the bounce-back fluid-structure
coupling with the multi-component fluid model of Shan and Chen.  By using a
specially designed reference model, the method has been shown to achieve an
excellent agreement for the deformation of a Skalak capsule at a fluid-fluid
interface, where the errors stemming from the coupling between the
multi-component fluid and the particle surface are well below the
characteristic length scale of the discretized fluid lattice.  Despite the
diffuse interface of the used multi-component model, a good agreement is still
observed for the equilibrium shapes of particles with radii comparable to the
diffuse interface width.

The used fluid-structure coupling furthermore allows one to easily tune the contact angle of a
single particle over a wide range of contact angles.  It has been shown that
the softness of the particle has a significant influence on the vertical
displacement of the center of mass of the particle relative to the fluid
interface, resulting from the additional deformation of the particle.

All individual simulations shown in this work can be comfortably run on a
modern desktop, leaving ample room to increase the system size and number of
modeled particles when using modern high-performance clusters.  The
implemented code has been successfully tested for large systems with up to 2000
particles on available high-performance computing clusters, and therefore paves
the way to investigate a variety of systems where both fluid-interfaces and the
softness of suspended particles play a significant role.

There are also multiple opportunities to straightforwardly increase the level
of complexity in the modeled systems within the presented method.  For
example, one can trivially move into the direction of inhomogeneous wetting
particle surfaces, i.e. soft Janus particles, by defining the particle color
individually on each boundary element.  Furthermore, the bounce-back coupling
can easily be altered to allow for a small permeability of the boundary by
applying the bounce-back coupling only to a fraction of the distribution
functions, allowing to model the slow release of the interior fluid of the soft
capsules.

The half-way bounce-back coupling introduces the well-known staircasing effect
in the discretization of the particle surface on the fluid lattice, thereby
adding small discretization errors in the normal vector of the particle
surface, which in turn result in small deviations of the expected equilibrium
position.  Furthermore, the bounce-back conditions introduce small force
fluctuations in the force acting on the surface when the discretization on the
fluid lattice changes (not shown in this paper).  If desired, these effects can
be mitigated by switching to higher order coupling schemes, such as
interpolated bounce-back, but one needs to keep in mind that such methods break
the local mass-conservation.  During the development of the present method the
small mass-errors from interpolated bounce-back schemes have been observed to
quickly result in a drift of the mass in the system, since the density
gradients close to the fluid interface are a priori large.  In contrast to the
mass change due to the fresh node interpolation, which was remedied by using
the adaptive rescaling proposed by Jansen et al., the mass change from
interpolated bounce-back also occurs for particles which are fixed in space and
time.  The used mass-correction term is unsuitable to correct for these errors,
and devising another mass-conserving scheme without significantly afflicting
the local fluid velocity is non-trivial.  Hence, for the applications of dense
suspensions the half-way bounce-back boundary condition is the most suitable,
since it only requires a single fluid node between two adjacent particles,
whereas the higher-order methods typically require two or three fluid nodes.

\begin{acknowledgments}
    Special thanks go out to Manuel Zellh{\"o}fer for help with various technical aspects during the code development.
    We acknowledge NWO/TTW (project 10018605), HPC-Europa3 (INFRAIA-2016-1-730897) and the DFG within the Cluster of Excellence "Engineering of Advanced Materials" (project EXC315, Bridge funding) for available funding, and the J{\"u}lich Supercomputing Centre and the High Performance Computing Centre Stuttgart for the allocated CPU time.
    TK thanks the University of Edinburgh for the award of a Chancellor's Fellowship.
\end{acknowledgments}

\end{document}